\documentclass[10pt,twocolumn,twoside] {IEEEtran} 

\usepackage{flushend}
\usepackage{setspace}
\usepackage[cmex10]{amsmath}
\usepackage{amsfonts}
\usepackage{amssymb}
\usepackage{algorithmic}
\usepackage{cite}
\usepackage{url}
\usepackage[framemethod=TikZ]{mdframed}
\usepackage{epstopdf}
\usepackage[font=footnotesize]{subfig}
\usepackage{graphicx}

\long\def\comment#1{}

\newfont{\bbb}{msbm10 scaled 700}

\newfont{\bb}{msbm10 scaled 1100}

\newcommand{\nv}{{\bf n}}

\newcommand{\qv}{{\bf q}}

\newcommand{\xv}{{\bf x}}
\newcommand{\yv}{{\bf y}}

\newcommand{\zerov}{{\bf 0}}

\newcommand{\Id}{{\bf I}}

\newcommand{\Km}{{\bf K}}

\newcommand{\Pm}{{\bf P}}

\newcommand{\Xm}{{\bf X}}

\newcommand{\Cc}{{\cal C}}

\newcommand{\Nc}{{\cal N}}
\newcommand{\Oc}{{\cal O}}

\newcommand{\Sc}{{\cal S}}

\newcommand{\phiv}{\hbox{\boldmath$\phi$}}

\newcommand{\Phim}{\hbox{\boldmath$\Phi$}}

\renewcommand{\det}{{\hbox{det}}}

\newcommand{\herm}{{\sf H}}

\newcommand{\transp}{{\sf T}}

\mdfdefinestyle{MyFrame}{%
    linecolor=black,
    outerlinewidth=1pt,
    roundcorner=10pt,
    innertopmargin=\baselineskip,
    innerbottommargin=\baselineskip,
    innerrightmargin=20pt,
    innerleftmargin=20pt,
    backgroundcolor=gray!5!white}

\begin{document}
\title{A Fast Non-Gaussian Bayesian Matching Pursuit Method for Sparse Reconstruction}

\author{Mudassir~Masood, and~Tareq Y. Al-Naf{}fouri*,~\IEEEmembership{Member,~IEEE,}
\thanks{Authors are with the Department
of Electrical Engineering, King Abdullah University of Science \& Technology, Thuwal 23955-6900, Kingdom of Saudi Arabia, 
e-mail: mudassir.masood@kaust.edu.sa, tareq.alnaffouri@kaust.edu.sa}
\thanks{Tareq Y. Al-Naf{}fouri is also associated with the Department of Electrical Engineering, King Fahd University of Petroleum and Minerals, Dhahran 31261, Kingdom of Saudi Arabia.}
\thanks{This work has been submitted to the IEEE for possible publication. Copyright may be transferred without notice, after which this version may no longer be accessible.}}

\maketitle

\begin{abstract}
A fast matching pursuit method using a Bayesian approach is introduced for sparse signal recovery. This method, referred to as nGpFBMP, performs Bayesian estimates of sparse signals even when the signal prior is non-Gaussian or unknown. It is agnostic on signal statistics and utilizes \textit{a priori} statistics of additive noise and the sparsity rate of the signal, which are shown to be easily estimated from data if not available. nGpFBMP utilizes a greedy approach and order-recursive updates of its metrics to find the most dominant sparse supports to determine the approximate minimum mean square error (MMSE) estimate of the sparse signal. Simulation results demonstrate the power and robustness of our proposed estimator. 
\end{abstract}

\begin{IEEEkeywords}
sparse reconstruction, compressed sensing, Bayesian, linear regression, matching pursuit, basis selection, minimum mean-square error (MMSE) estimate, greedy algorithm
\end{IEEEkeywords}

\section{Introduction}
\IEEEPARstart{S}{parsity} is a feature that is abundant in both natural and man-made signals. Some examples of sparse signals include those from speech, images, videos, sensor arrays (e.g., temperature and light sensors), seismic activity, galactic activities, biometric activity, radiology, and frequency hopping. Given the vast existence of signals, their sparsity is an attractive property because the exploitation of this sparsity may be useful in the development of simple signal processing systems. Some examples of systems in which \textit{a priori} knowledge of signal sparsity is utilized include motion estimation \cite{6126522}, magnetic resonance imaging (MRI) \cite{4472246},  impulse noise estimation and cancellation in DSL \cite{5938020}, network tomography\cite{5684036}, and peak-to-average-power ratio reduction in OFDM \cite{patent:20110122930}. All of these systems are based on sparsity-aware estimators such as Lasso \cite{ISI:A1996TU31400017}, basis pursuit \cite{ISI:000075434800003}, structure-based estimator \cite{6120179}, fast Bayesian matching pursuit \cite{FBMP}, and estimators related to the relatively new area of compressed sensing \cite{ISI:000246789100008, ISI:000234944700009,ISI:000236714000001}.

Compressed sensing (CS), otherwise known as compressive sampling, has found many applications in the fields of communications, image processing, medical imaging, and networking. CS algorithms have been shown to recover sparse signals from underdetermined systems of equations that take the form 
\begin{equation}
\yv = \Phim \xv +\nv
\end{equation}
where $\xv \in \mathbb{C}^N$, and $\yv \in \mathbb{C}^M$ are the unknown sparse signal and the observed signal, respectively. Furthermore, $\Phim \in \mathbb{C}^{M\times N}$ is the measurement matrix and $\nv \in \mathbb{C}^M$ is the additive Gaussian noise vector.  Here, the number of unknown elements, $N$, is much larger than the number of observations, $M$. CS uses linear projections of sparse signals that preserve structure of signals; furthermore, these projections are used to reconstruct the sparse signal using $l_1$-optimization with high probability.
\begin{equation}
\hat{\xv} = \mathbf{argmin} \left\|\xv\right\|_1 \text{ such that } \Phim \xv = \yv
\end{equation}
$l_1$-optimization is a convex optimization problem that conveniently reduces to a linear program known as basis pursuit, which has the computational complexity of $\mathcal{O}(N^3)$. Since, usually, $N$ is large, such an approach rapidly becomes unrealistic. To tackle this problem, some efficient alternatives such as orthogonal matching pursuit (OMP) \cite{ ISI:000251801500016} and the algorithm proposed by Haupt et al. \cite{ISI:000240076700011} have been proposed. These algorithms fall into the category of greedy algorithms that are relatively faster than basis pursuit. However, an inherent problem in these systems is that the only \textit{a priori} information utilized is the sparsity information.

Another category of methods based on the Bayesian approach considers complete \textit{a priori} statistical information of sparse signals. A method called fast Bayesian matching pursuit (FBMP) \cite{FBMP}, adopts such an approach and assumes a Gaussian prior on the unknown sparse vector, $\xv$. This method performs sparse signal estimation via model selection and model averaging. The sparse vector is described as a mixture of several components, the selection of which is based on successive model expansion. FBMP obtains an approximate MMSE estimate of the unknown signal with high accuracy and low complexity. It was shown to outperform several sparse recovery algorithms, including OMP \cite{ISI:000251801500016}, StOMP \cite{ ISI:000300246900039}, GPSR-Basic \cite{ISI:000265494900006}, Sparse Bayes \cite{ ISI:000173336900003}, BCS \cite{Ji:2007:BCS:1273496.1273544} and a variational-Bayes implementation of BCS \cite{Bishop:2000:VRV:2073946.2073953}. However, there are several drawbacks associated with this method. The assumption that the signal prior is Gaussian is not realistic, because, in most real-world scenarios, the signal distribution is not Gaussian, or it is unknown. In addition, its performance is dependent on the knowledge of signal statistics, which are usually difficult to compute. Although, a signal statistics estimation process is proposed, it is dependent on knowledge of the initial estimates of these signal parameters. The estimation process, in turn, has a negative impact on the complexity of the method.

Another popular Bayesian method proposed by Larsson and Sel\'{e}n \cite{ISI:000243952600005} computes the MMSE estimate of the unknown vector, $\xv$. Its approach is similar to that of FBMP in the sense that the sparse vector is described as a mixture of several components that are selected based on successive model reduction. It also requires knowledge of the noise and signal statistics. However, it was found that the MMSE estimate is insensitive to the \textit{a priori} parameters and therefore an empirical-Bayesian variant that does not require any \textit{a priori} knowledge of the data was devised.

The Bayesian approaches mentioned above work successfully only for Gaussian priors. It is reasonable to assume that any additive noise, generated at the sensing end, is Gaussian. However, assuming the signal statistics to be Gaussian is inadequate, because the actual situation is not captured. Moreover, in situations where the assumption of a Gaussian prior is valid, the parameters of that prior (mean and covariance) need to be estimated, which is challenging, especially when the signal statistics are not i.i.d. In that respect, one can appreciate convex relaxation approaches that are agnostic with regard to signal statistics.

In this paper, we pursue a Bayesian approach for sparse signal reconstruction that combines the advantages of the two approaches. On one hand, the approach is Bayesian, acknowledging the noise statistics and the signal sparsity rate, while on the other hand, the approach is agnostic on the signal amplitude statistics. The approach can bootstrap itself and estimate the required statistics (sparsity rate and noise variance) when they are unknown. The algorithm is implemented in a greedy manner and pursues an order-recursive approach, helping it to enjoy low complexity. Specifically, the advantages of our approach are as follows
\begin{enumerate}
\item The Bayesian estimate of the sparse signal is performed even when the signal prior is non-Gaussian or unknown.
\item Signals whose statistics are unknown are dealt with. In fact, contrary to other methods, these statistics need not be estimated, which is particularly useful when the signal statistics are not i.i.d. Therefore, it is agnostic with regard to variations in signal statistics.
\item The \textit{a priori} statistics of the additive noise and the sparsity rate of the signal, which can be easily estimated from the data if not available, are utilized.
\item The greedy nature of the approach and the order-recursive update of its metrics, make it simple.
\end{enumerate}

The remainder of this paper is organized as follows. In Section \ref{sec:problemformulation}, we formulate the problem and present the MMSE setup in the non-Gaussian/unknown statistics case. In Section \ref{sec:greedy}, we describe our greedy algorithm that is able to obtain the approximate MMSE estimate of the sparse vector. Section \ref{sec:efficientdssm} demonstrates how the greedy algorithm can be made faster by calculating various metrics in a recursive manner. This is followed by Section \ref{sec:estimation}, which describes our hyperparameter estimation process. In Section \ref{sec:results}, we present our simulation results and in Section \ref{sec:conclusions}, we conclude the paper.

\subsection{Notation}
We denote scalars with small-case letters (e.g., $x$), vectors with small-case, bold-face letters (e.g., $\xv$), matrices with upper-case, bold-face letters (e.g., $\Xm$), and we reserve calligraphic notation (e.g., $\Sc$) for sets. We use $\xv_i$ to denote the $i^{th}$ column of matrix $\Xm$, $x(j)$ to denote the $j^{th}$ entry of vector $\xv$, and $\Sc_i$ to denote a subset of a set $\Sc$. We also use $\Xm_{\Sc}$ to denote the sub-matrix formed by the columns $\{\xv_i : i \in \Sc \}$, indexed by set $\Sc$. Finally, we use $\hat{\xv}$, $\xv^*$, $\xv^\transp$, and $\xv^\herm$ to respectively denote the estimate, conjugate, transpose, and conjugate transpose of the vector $\xv$.

\section{Problem Formulation and MMSE Setup}\label{sec:problemformulation}
\subsection{The Signal Model}

The analysis in this paper considers obtaining an $N\times 1$ sparse vector, $\xv$, from an $M\times 1$ observations vector, $\yv$. These observations obey the linear regression model
\begin{equation}\label{eq:sigmodel}
\yv = \Phim\xv + \nv
\end{equation}
where $\Phim$ is a known $M\times N$ regression matrix and $\nv\sim\Cc\Nc(\zerov, \Km_\nv)$ is the additive Gaussian noise vector.

We shall assume that $\xv$ has a sparse structure and is modeled as $\xv =\xv_A \circ \xv_B$ where $\circ$ indicates Hadamard (element-by-element) multiplication. The vector $\xv_A$ consists of elements that are drawn from some unknown distribution and the entries of $\xv_B$ are drawn i.i.d. from a Bernoulli distribution with success probability $p$. We observe that the sparsity of vector $\xv$ is controlled by $p$ and, therefore, we call it the sparsity parameter/rate. Typically, in Bayesian estimation, the signal entries are assumed to be drawn from a Gaussian distribution but here we would like to emphasize that the distribution and color of the entries of $\xv_A$ do not matter.\footnote{The distribution may be unknown or known with unknown parameters or even Gaussian. Our developments are agnostic with regard to signal statistics.}

\subsection{MMSE Estimation of $\xv$}

To determine $\xv$, we compute the MMSE estimate of $\xv$ given observation $\yv$. This estimate is formally defined by

\begin{equation}\label{eq:xmmse}
\hat{\xv}_{mmse} \triangleq \mathbb{E}[\xv|\yv] = \sum_{\Sc} p(\Sc|\yv)\mathbb{E}[\xv|\yv,\Sc]
\end{equation}

\noindent where the sum is executed over all possible $2^N$ support sets of $\xv$. In the following, we explain how the expectation $\mathbb{E}[\xv|\yv,\Sc]$, the posterior $p(\Sc|\yv)$ and the sum in (\ref{eq:xmmse}) can be evaluated.

Given the support $\Sc$, (\ref{eq:sigmodel}) becomes 
\begin{equation}
\yv = \Phim_\Sc \xv_\Sc+ \nv
\end{equation}
where $\Phim_\Sc$ is a matrix formed by selecting columns of $\Phim$ indexed by support $\Sc$. Similarly, $\xv_\Sc$ is formed by selecting entries of $\xv$ indexed by $\Sc$. Since the distribution of $\xv$ is unknown, the best we can do is to use the best linear unbiased estimate (BLUE) as an estimate of $\xv$, i.e.,
\begin{equation}\label{eq:blue}
\mathbb{E}[\xv|\yv,\Sc] = \left( \Phim_\Sc^\herm \Phim_\Sc \right)^{-1} \Phim_\Sc^\herm \yv
\end{equation}
The posterior in (\ref{eq:xmmse}) can be written using the Bayes rule as

\begin{equation}\label{eq:posteriors}
p(\Sc|\yv) = \frac{p(\yv|\Sc) p(\Sc)}{p(\yv)}
\end{equation}
The probability, $p(\yv)$, is a factor common to all posterior probabilities that appear in \ref{eq:sigmodel} and hence can be ignored. Since the elements of $\xv$ are activated according to the Bernoulli distribution with success probability $p$, we have
\begin{equation}\label{eq:ps}
p(\Sc) = p^{|\Sc|} (1-p)^{N-|\Sc|}
\end{equation}
It remains to evaluate the likelihood $p(\yv|\Sc)$. If $\xv_\Sc$ is Gaussian, $p(\yv|\Sc)$ would also be Gaussian and that is easy to evaluate. On the other hand, when the distribution of $\xv$ is unknown or even when it is known but non-Gaussian, determining $p(\yv|\Sc)$ is in general very difficult. To go around this, we note that $\yv$ is formed by a vector in the subspace spanned by the columns of $\Phim_\Sc$ plus a Gaussian noise vector, $\nv$. This motivates us to eliminate the non-Gaussian component by projecting $\yv$ onto the orthogonal complement space of $\Phim_\Sc$. This is done by multiplying $\yv$ by the projection matrix $\Pm_\Sc^\bot =  \Id - \Phim_\Sc\left( \Phim_\Sc^\herm \Phim_\Sc\right)^{-1} \Phim_\Sc^\herm$. This leaves us with $\Pm_\Sc^{\bot} \yv = \Pm_\Sc^{\bot} \nv$, which is Gaussian with a zero mean and covariance 
\begin{align}
\mathbf{K} &= \mathbb{E}[(\Pm_\Sc^{\bot} \nv)(\Pm_\Sc^{\bot} \nv)^\herm]\nonumber\\
&= \Pm_\Sc^{\bot} \mathbb{E}[\nv\nv^\herm] {\Pm_\Sc^{\bot}}^\herm\nonumber\\
&= \Pm_\Sc^{\bot} \Km_\nv {\Pm_\Sc^{\bot}}^\herm
\end{align}
where $\Km_\nv$ is the noise covariance matrix. Thus,
\begin{equation} 
p(\yv|\Sc) \simeq \frac{1}{\sqrt{(2\pi)^M\det \mathbf{K}}}\exp\left(-\frac{1}{2}\left(\Pm_\Sc^{\bot} \yv\right)^\herm \mathbf{K}^{-1} \left(\Pm_\Sc^{\bot} \yv\right)\right)
\end{equation}
Dropping the pre-exponential factor yields
\begin{equation}
p(\yv|\Sc) \propto \exp\left(-\left\| \Pm_\Sc^{\bot} \yv \right\|_{\Km^{-1}}^2\right)
\end{equation}
In the white noise case, $\nv \sim \Nc(\zerov, \sigma_\nv^2\Id)$, which we will focus on for the remainder of the paper, we have
\begin{equation}\label{eq:pys}
p(\yv|\Sc) \simeq \exp\left(-\frac{1}{\sigma_\nv^2}\left\| \Pm_\Sc^{\bot} \yv \right\|^2\right)
\end{equation}
Substituting (\ref{eq:ps}) and (\ref{eq:pys}) into (\ref{eq:posteriors}) finally yields an expression for the posterior probability. In this way, we have all the ingredients to compute the sum in (\ref{eq:xmmse}). Computing this sum is a challenging task when $N$ is large because the number of support sets can be extremely large and the computational complexity can become unrealistic. To have a computationally feasible solution, this sum can be computed over a few support sets corresponding to significant posteriors. Let $\Sc^d$ be the set of supports for which the posteriors are significant. Hence, we arrive at the following approximation to the MMSE estimate
\begin{equation}\label{eq:xammse}
\hat{\xv}_{ammse} = \sum_{\Sc^d} p(\Sc^d|\yv)\mathbb{E}[\xv|\yv,\Sc^d]
\end{equation}
In the next section, we propose a greedy algorithm to find $\Sc^d$. Before proceeding, for ease of representation and convenience, we represent the posteriors in the log domain. In this regard, we define a dominant support selection metric, $\nu(\Sc)$, to be used by the greedy algorithm as follows:
\begin{align}
\nu(\Sc) &\triangleq \ln p(\yv|\Sc) p(\Sc)\nonumber\\
&=\ln {} \exp \left( -\frac{1}{\sigma_\nv^2} \left\|\Pm_\Sc^{\bot} \yv\right\|^2 \right) + \ln \left(p^{|\Sc|} (1-p)^{N-|\Sc|}\right)\nonumber\\
&=  \frac{1}{\sigma_\nv^2} \left\|\Phim_\Sc (\Phim_\Sc^\herm \Phim_\Sc)^{-1}\Phim_\Sc^\herm \yv\right\|^2   -\frac{1}{\sigma_\nv^2}  \left\| \yv\right\|^2 \nonumber\\
&\qquad\qquad+   \left|\Sc\right| \ln p  + (N-\left|\Sc\right|) \ln (1-p) \label{eq:dssm}
\end{align}

\section{A Greedy Algorithm}\label{sec:greedy}

\begin{figure}
	\centering
		\includegraphics[scale=0.32]{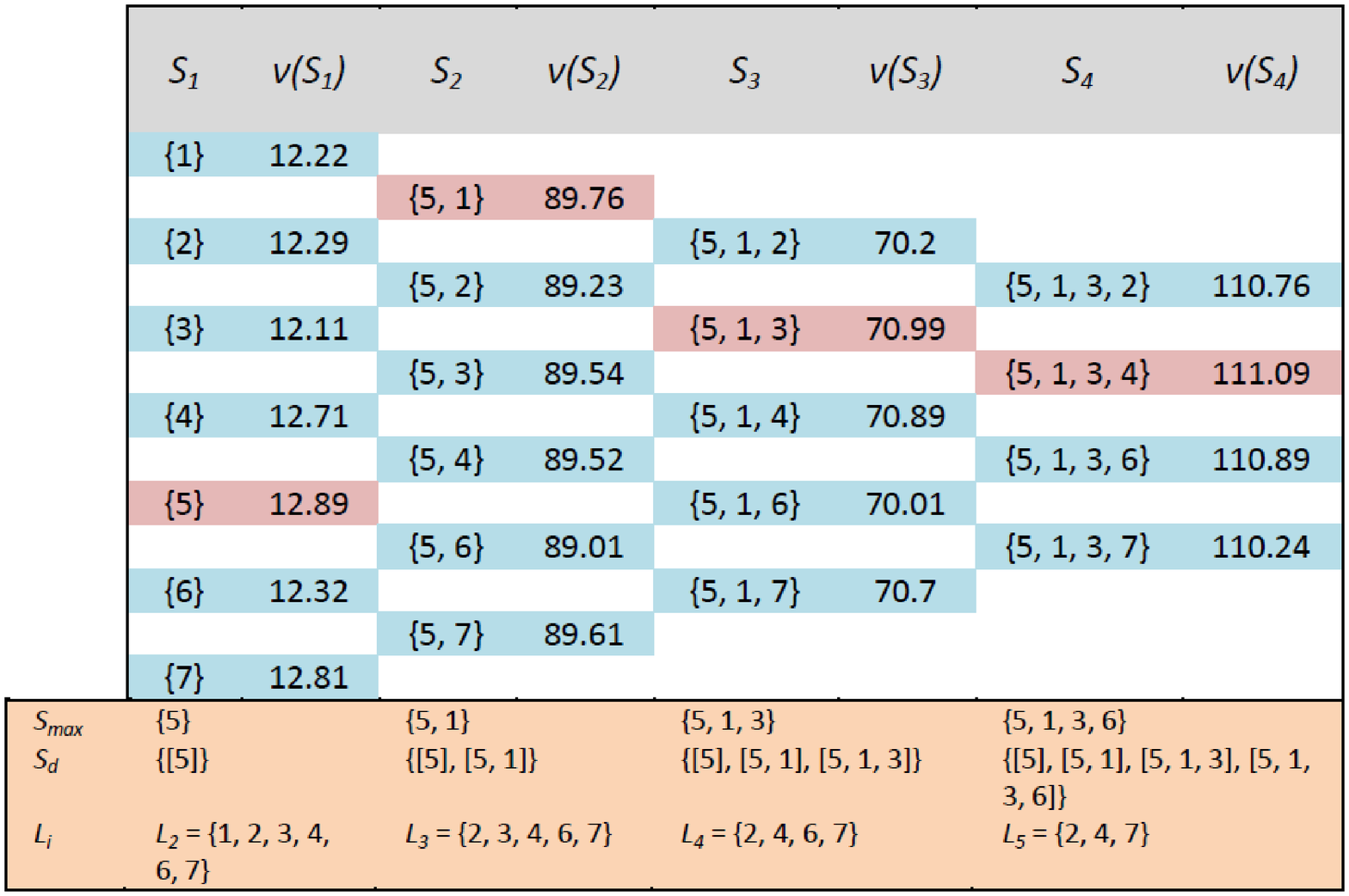}
	\caption{An example run of the greedy algorithm for $N=7$ and $P=4$}
	\label{fig:algoexample}
\end{figure}

We now present a greedy algorithm to determine the set of dominant supports, $\Sc^d$, required to evaluate $\hat{\xv}_{ammse}$ (\ref{eq:xammse}). We search for the optimal support in a greedy manner. We first start by finding the best support of size 1, which involves evaluating $\nu(\Sc)$ for $\Sc=\{ 1 \}, \dots, \{ N \}$, i.e., a total of $\binom{N}{1}$ search points. Let $\Sc_1 = \{ i_1^\star \}$ be the optimal support. Now, we look for the optimal support of size 2, which involves a search of size $\binom{N}{2}$. To reduce the search space, we pursue a greedy approach and look for the point $i_2^\star \neq i_1^\star$ such that $\Sc_2=\{ i_1^\star, i_2^\star \}$ maximizes $\nu(\Sc_2)$. This involves $\binom{N-1}{1}$ search points (as opposed to the optimal search over $\binom{N}{2}$ points). We continue in this manner by forming $\Sc_3 = \{ i_1^\star, i_2^\star, i_3^\star \}$ and searching for $i_3^\star$ in the remaining $N-2$ points and so on until we reach $\Sc_P =\{  i_1^\star, \dots, i_P^\star\}$. The value of $P$ is selected to be slightly larger than the expected number of nonzero elements in the constructed signal such that $\Pr(|\Sc|>P)$ is sufficiently small\footnote{$|\Sc|$, i.e., support of the constructed signal, follows the binomial distribution $\mathcal{B}(N,p)$, which can be approximated by the Gaussian distribution $\mathcal{N}(Np, Np(1-p))$ if $Np>5$. For this case, $\Pr(|\Sc| > P) = \frac{1}{2}\text{erfc}{\frac{P-Np}{\sqrt{2Np(1-p)}}}$.}\label{ft:calculateP}. A formal algorithmic description is presented in Table \ref{alg:greedy} and an example run of this algorithm for $N=7$ and $P=4$ is presented in Fig. \ref{fig:algoexample}.

One point to note here is that in our greedy move from $\Sc_j$ to $\Sc_{j+1}$, we need to evaluate $\nu(\Sc_j \cup \{i_{j+1}\})$ around $N$ times, which can be done in an order-recursive manner starting from that of $\nu(\Sc_j)$. Specifically, we note that every expansion, $\Sc_j \cup \{i_{j+1}\}$, from $\Sc_j$ requires a calculation of $\nu(\Sc_j \cup \{i_{j+1}\})$ from (\ref{eq:dssm}). This translates to appending a column, $\phiv_{j+1}$, to $\Phim_{\Sc_j}$ in the calculations of (\ref{eq:dssm}), which can be done in an order-recursive manner. We summarize these calculations in Section \ref{sec:efficientdssm}. This order-recursive approach reduces the calculation in each search step to an order of $\Oc(MN)$ operations down from $\Oc(MN^2)$ in the direct approach. Therefore, the complexity we incur is of the order $\Oc(PMN)$ in our greedy search for the best $P$ support.

\begin{table}
\caption{The Greedy Algorithm}
\label{alg:greedy}
\begin{mdframed}
\begin{enumerate}
\item Initialize $L=\{1, 2, \hdots, N\}$, $\Sc_{max}=\{\}$, $\Sc^d=\{\}$, $i=1$, $L_i = L$.
\item If $i>P$, then stop.\label{itm:gs1}
\item Generate $\Omega =  \{ \Sc_{max} \cup \{ \alpha_1 \},
 \Sc_{max} \cup \{ \alpha_2 \},
\cdots,
 \Sc_{max} \cup \{ \alpha_{|L_i|} \} \mid \alpha_k \in L_i\}$
\item Compute $\{ \nu(\Sc_k) \mid \Sc_k \in \Omega \}$.\label{itm:nu}
\item Find $\Sc^\star \in \Omega$, such that $\nu(\Sc^\star) \ge \max_j \nu(\Sc_j)$.
\item Update, $\Sc^d = \{\Sc^d, \Sc^\star\}$, $\Sc_{max}=\Sc^\star$, $L_{i+1}= L~\backslash~\Sc^\star$.
\item Set $i \leftarrow i+1$ and repeat steps \ref{itm:gs1} - \ref{itm:gslast}. \label{itm:gslast}
\end{enumerate}
\end{mdframed}
\end{table}

\subsection{A Repeated Greedy Search}
It is obvious that to achieve the best signal estimate, $\Sc^d$ should contain all $2^N$ possibilities of supports. However, our greedy algorithm would result in an approximation of the signal estimate because it selects only a handful of supports. The accuracy of the reconstructed signal is dependent on the number of support vectors in $\Sc^d$ and may be increased by repeating the greedy algorithm a number of times (e.g., $D$). This would result in $\Sc^d$ with $D$ number of support vectors for each of the $P$ sparsity levels (i.e., a total of $PD$ supports). The selection of supports in subsequent repetitions of the algorithm is performed by making sure not to select an element at a particular sparsity level that has been selected at the same sparsity level in any of the previous repetitions. We note that a repeated greedy search in this manner would incur a complexity of order $\Oc(DPMN)$.  For a detailed description of the steps followed by the method, the pseudocode is provided in the Appendix and the nGpFBMP code is provided on the author's website\footnote{The MATLAB code of the nGpFBMP algorithm and the results from various experiments discussed in this paper are provided at http://faculty.kfupm.edu.sa/ee/naffouri/publications.html}.
\subsection*{Remark}
Let $\Sc_j^{(1)}$ and $\Sc_j^{(2)}$ be two different support vectors at the $j$th sparsity level. Note from (\ref{eq:dssm}) that if $\nu(\Sc_j^{(1)}) > \nu(\Sc_j^{(2)})$ for some $\sigma_\nv^2$ and $p$, then that inequality remains valid regardless of how $\sigma_\nv^2$ and $p$ change. In other words, the selection of dominant supports at each sparsity level is independent of these quantities. This observation helps the algorithm to bootstrap itself and to estimate the unknown sparsity rate and noise variance, as demonstrated ahead.

\section{Efficient Computation of the Dominant Support Selection Metric}\label{sec:efficientdssm}
As explained in Section \ref{sec:greedy}, $\nu(\Sc)$ requires extensive computation to determine the dominant supports. The computational complexity of the proposed algorithm is therefore largely dependent upon the way $\nu(\Sc)$ is computed. In this section, a computationally efficient procedure to calculate this quantity is presented.

We note that the efficient computation of $\nu(\Sc)$ depends mainly on the efficient computation of the term $\mathbf{\xi}_\Sc = \left\|\left( \Phim_\Sc (\Phim_\Sc^\herm \Phim_\Sc)^{-1}\Phim_\Sc^\herm \yv\right)  \right\|^2 =  \left\|\Phim_\Sc \mathbb{E}[\xv|\yv,\Sc]\right\|^2$. Our focus is therefore on computing $\mathbb{E}[\xv|\yv,\Sc]$ efficiently. 

Consider the general support $\Sc = \{ s_1, s_2, s_3, \dots, s_k \}$ with $s_1 < s_2 < \dots < s_k$ and let $\underline{\Sc}$ and $\overline{\Sc}$ denote the following subset and superset, respectively:
\begin{align}
\underline{\Sc} = \{ s_1, s_2, s_3, \dots,s_{k-1} \} \quad
\overline{\Sc} = \{ s_1, s_2, s_3, \dots,s_{k+1} \}
\end{align}
where $s_k <s_{k+1}$. In the following, we demonstrate how to update $\mathbf e_{\yv, k-1} (\underline{\Sc}) \triangleq \mathbb{E} [\xv_{\underline{\Sc}} | \, \yv]$ to obtain\footnote{We explicitly indicate the size $k$ of $\Sc$ in this notation as it elucidates the recursive nature of the developed algorithms.} $\mathbf e_{\yv, k}(\Sc)=\mathbb{E}[\xv_{\Sc} | \, \yv]$. Here, we use $\Sc$ to designate the supports and thus $\mathbb{E}[\xv_{\Sc} | \, \yv]$ refers to $\mathbb{E}[\xv|\yv,\Sc]$. We note that
\begin{align}
\mathbf {e}_{\yv,k}{(\Sc)} &= \left( \boldsymbol \Phi_{\Sc}^\herm \boldsymbol \Phi_{\Sc} \right)^{-1} \boldsymbol \Phi_{\Sc}^\herm \yv = \left( \left[
\begin{matrix}
\boldsymbol \Phi_{\underline{\Sc}}^\herm \\
\\
 \boldsymbol \phi_{s_k}^\herm
\end{matrix}
\right]
\left[ \boldsymbol \Phi_{\underline{\Sc}} \boldsymbol \phi_{s_k} \right] \right)^{-1}   \left[
\begin{matrix}
\boldsymbol \Phi_{\underline{\Sc}}^\herm  \yv \\
\\
\boldsymbol \phi_{s_k}^\herm  \yv
\end{matrix}
\right]
\end{align}
By using the block inversion formula to express the inverse of the above and simplifying, we get

\begin{align}
\label{eq:ey_recur}
\mathbf {e}_{\yv, k}(\Sc)
&=
\left[
\begin{array}{ll}
&\frac{1}{f_\Sc}  \left( \qv_{\boldsymbol {\phi}, k}^\herm ({\Sc}) \mathbf {e}_{\yv, k-1}{(\underline{\Sc})} -  \mathbf {e}_{\yv, 1}(s_k) \right) \mathbf {e}_{\boldsymbol \phi, k}(\Sc) \\
&\qquad\qquad+ \mathbf {e}_{\yv, k-1}(\underline {\Sc}) \\
\\
&\frac{-1}{f_\Sc} \left( \qv_{\boldsymbol {\phi}, k}^\herm ({\Sc}) \mathbf {e}_{\yv, k-1}(\underline {\Sc}) -  \mathbf {e}_{\yv, 1}(s_k) \right)
\end{array}
\right]
\end{align}
This recursion is initialized by $\mathbf {e}_{\yv, 1}(i) = \left(\boldsymbol \phi_{s}^\herm \boldsymbol \phi_{s}\right)^{-1} \boldsymbol \phi_{s}^\herm  \yv$ for $i=1, 2, \dots, N$. The recursion also depends on $\qv_{\boldsymbol \phi, k}({\Sc}) \triangleq \boldsymbol \Phi_{\underline{\Sc}}^\herm \boldsymbol \phi_{s_k}$, $\mathbf {e}_{\boldsymbol \phi, k}({\Sc}) \triangleq (\boldsymbol \Phi_{\underline{\Sc}}^\herm \boldsymbol \Phi_{\underline {\Sc}})^{-1} \boldsymbol \Phi_{\underline{\Sc}}^\herm \boldsymbol \phi_{s_k}$ and $f_\Sc \triangleq 1 - \qv_{\boldsymbol {\phi}, k}^\herm ({\Sc}) \mathbf {e}_{\boldsymbol \phi, k}(\Sc)$. The recursions for $\qv_{\boldsymbol \phi, k}({\Sc})$, and $\mathbf {e}_{\boldsymbol \phi, k}({\Sc})$ may be determined in a similar fashion as
\begin{align}
\label{eq:e_recurs}
\mathbf {e}_{\boldsymbol \phi, k+1}(\overline {\Sc})
&=
\left[
\begin{array}{ll}
\frac{1}{f_\Sc} \left( \qv_{\boldsymbol {\phi}, k}^\herm ({\Sc}) \mathbf {e}_{\boldsymbol \phi, k}(\underline{\Sc};s_{k+1}) - \mathbf {e}_{\boldsymbol \phi, 2}(s_k;s_{k+1}) \right) \\
\qquad \qquad \mathbf {e}_{\boldsymbol \phi, k}(\Sc) + \mathbf {e}_{\boldsymbol \phi, k}(\underline {\Sc};s_{k+1})\\
\\
\frac{-1}{f_\Sc} \left( \qv_{\boldsymbol {\phi}, k}^\herm ({\Sc}) \mathbf {e}_{\boldsymbol \phi, k}(\underline{\Sc};s_{k+1}) - \mathbf {e}_{\boldsymbol \phi, 2}(s_k;s_{k+1}) \right)
\end{array}
\right]
\end{align}
and\footnote{Notation such as $\mathbf {e}_{\boldsymbol \phi, k}(\underline{\Sc};s_{k+1})$ is a short hand for $\mathbf {e}_{\boldsymbol \phi, k}(\underline{\Sc} \cup \{s_{k+1}\})$.}
\begin{align}
\label{eq:q_recurs}
\qv_{\boldsymbol {\phi}, k+1} (\overline{\Sc}) &=
\left[
\begin{matrix}
\boldsymbol \Phi_{\underline {\Sc}}^\herm \\
\boldsymbol \phi_{s_k}^\herm
\end{matrix}
\right]
\boldsymbol \phi_{s_{k+1}}
=
\left[
\begin{matrix}
\qv_{\boldsymbol {\phi}, k} (\underline {\Sc};s_{k+1}) \\
\qv_{\boldsymbol {\phi}, 2} (s_{k};s_{k+1})
\end{matrix}
\right]
\end{align}
The two recursions (\ref{eq:e_recurs}) and (\ref{eq:q_recurs}) start at $k=2$ and are thus initialized by $\mathbf {e}_{\boldsymbol \phi, 2}(s_1; s_2)$ and $\qv_{\boldsymbol {\phi}, 2} (s_1; s_2)$ for $s_1, s_2 = 1, 2, \dots, N$. This completes the recursion of $\mathbf {e}_{\yv,k}{(\Sc)}$ which we utilize for recursive evaluation of $\nu(\Sc)$ as
\begin{align}
\nu_k(\Sc) &=  \frac{1}{\sigma_\nv^2} \left\|\Phim_\Sc  \mathbf {e}_{\yv, k}(\Sc)\right\|^2   -\frac{1}{\sigma_\nv^2}  \left\| \yv\right\|^2 +   \left|\Sc\right| \ln p  \nonumber\\
&\qquad\qquad+ (N-\left|\Sc\right|) \ln (1-p)
\end{align}

\section{Estimation of the hyperparameters $p$ and $\sigma_\nv^2$}\label{sec:estimation}

One of the advantages of the proposed nGpFBMP is that it is agnostic with regard to signal statistics; the only parameters required are the noise variance, $\sigma_\nv^2$, and the sparsity rate, $p$. In Section \ref{sec:greedy}, we pointed out that the dominant support selection process is independent of parameters $p$ and $\sigma_\nv^2$. We are therefore able to determine the supports irrespective of the initial estimates of these parameters. The independence of the dominant support selection process from $p$ and $\sigma_\nv^2$ allows accurate and rapid estimation of these parameters. We note that although $p$ and $\sigma_\nv^2$ are not needed for support calculations, they are required for computing the posteriors used in the calculation of $\hat{\xv}_{ammse}$ in (\ref{eq:xammse}). To determine these estimates, we might opt for finding the maximum-likelihood (ML) or maximum \textit{a posteriori} (MAP) estimates using the expectation maximization (EM) algorithm. However, this will add to the computational complexity and is unnecessary as a fairly accurate estimation could be performed in a very simple manner as follows. 

The maximum \textit{a posteriori} (MAP) estimate of support $\Sc$ is given by $\hat{\Sc}_{map} = \text{arg\,max}_{\Sc} p(\Sc|\yv)$. We use this to get the MAP estimate of $\xv$, i.e., $\hat{\xv}_{map} = \mathbb{E}[\xv|\yv,\hat{\Sc}_{map}]$. This $\hat{\xv}_{map}$ is in turn used to estimate $p$, iteratively, as follows:
\begin{align} \label{eq:phat}
\hat{p}^{(i+1)} = \left\|\hat{\xv}_{map}^{(i)}\right\|_0/N
\end{align}
Here, the superscripts refer to a particular iteration. The estimate is computed iteratively where in the first iteration of nGpFBMP, $\hat{p}^{(1)}$ is initialized by $p_{\textbf{init}}$, the given initial estimate, to compute $\hat{\xv}_{map}^{(1)}$. This is used to find the new estimate, $\hat{p}^{(2)}$, using (\ref{eq:phat}) which is then supplied to nGpFBMP in the next iteration to compute $\hat{\xv}_{map}^{(2)}$. This process is repeated until the estimate of $p$ changes by less than $2\%$ or until a predetermined maximum number of iterations has been performed. Simulation results show that, in most cases, $\hat{p}$ converges rapidly. At this stage, the estimate of the noise variance can be computed as follows:
\begin{align} 
\hat{\sigma}_n^2 = \text{var} (\yv-\mathbf{\Phi} \hat{\xv}_{map})
\end{align}
We note that we do not need any iteration for estimating the noise variance. We use the $\hat{\xv}_{map}$ corresponding to the final estimate, $\hat{p}$, to find $\hat{\sigma}_n^2$.

\section{Results}\label{sec:results}
To demonstrate the performance of the proposed nGpFBMP, we compare it here with Fast Bayesian Matching Pursuit (FBMP)\cite{FBMP} and the convex relaxation-based ($l_1$) approach. The reason FBMP was selected is that it was shown to outperform a number of the contemporary sparse signal recovery algorithms, including OMP \cite{ISI:000251801500016}, StOMP \cite{ ISI:000300246900039}, GPSR-Basic \cite{ISI:000265494900006}, and BCS \cite{Ji:2007:BCS:1273496.1273544}. Comparison with FBMP shows that nGpFBMP performs where FBMP fails for various signal settings which are discussed in detail in the following.
\subsection*{Signal Setup}
Experiments were conducted for signals drawn from Gaussian as well as non-Gaussian distributions. The following signal configurations were used for the experiments:
\begin{enumerate}
\item Gaussian (i.i.d. ($\mu_\xv = 10, \quad \sigma_\xv^2 = 2$)).
\item Non-Gaussian (Uniform, non-i.i.d. ($ 5 \le \mu_\xv \le 10, \quad 1 \le \sigma_\xv^2 \le 2$)).
\item Unknown distribution (for this example, different images with unknown statistics were used),
\end{enumerate}
where $\mu_\xv$ and $\sigma_\xv^2$ refer to the mean and variance of the corresponding distributions, respectively.

Entries of $M \times N$ sensing/measurement matrix $\mathbf{\Phi}$ were i.i.d., with zero means and complex Gaussian distribution where the columns were normalized to the unit norm. The size of $\mathbf{\Phi}$ selected for the experiments was $M=256, N=1024$. The noise had a zero mean and was white and Gaussian, $\mathcal{CN}(\mathbf{0}, \sigma_\nv^2 \mathbf{I}_M)$, with $\sigma_\nv^2$ determined according to the desired signal-to-noise ratio (SNR). Initial estimates of the hyperparameters used for the simulations were ${\mu_\xv}_\textbf{ est}=0$, ${\sigma_\xv^2}_\textbf{ est} = \frac{1}{10} \times\sigma_\xv^2$, ${\sigma_\nv^2}_\textbf{ est} = 10 \times\sigma_\nv^2$, and $p_\textbf{est}=0.003$, where estimates of the signal mean and variance were needed for FBMP.

In all of the experiments, parameter refinement was performed for both nGpFBMP and FBMP. For FBMP, the surrogate EM method proposed by its authors was used to refine the hyperparameters. The refinements were allowed to perform for a maximum of $E_{max}=10$ iterations. For fairness, support and amplitude refinement \cite{5938020} procedures were performed on the results of the CS algorithm\footnote{Actual parameter values were provided to the CS algorithm instead of estimates; furthermore, support and amplitude refinement was also performed to demonstrate that, despite these measures, its performance was inferior to that of nGpFBMP.}. Finally, the normalized mean-squared error (NMSE) between the original signal, $\xv$, and its MMSE estimate, $\hat{\xv}_{ammse}$, was used as the performance measure:
\begin{equation}
\epsilon = 10 \log_{10} \left( \frac{1}{K} \sum_{k=1}^K \frac{\left\| \hat{\xv}_k - \xv_k \right\|^2}{\left\| \xv_k \right\|^2} \right)
\end{equation}
where $K$ was the number of trials performed to compute NMSE between $\xv$ and its estimate. 

\subsection*{Experiment 1 (Signal estimation performance comparison for varying SNR)}

In the first set of experiments, NMSEs were measured for values of SNR between 0 dB and 30 dB and plotted to compare the performance of nGpFBMP with FBMP and the CS algorithm. The signal sparsity rate selected for these experiments was $p=0.005$.

\begin{figure*}[htbp]
\centerline{\subfloat[NMSE vs SNR for uniform non-i.i.d. input]{\includegraphics[scale=0.5]{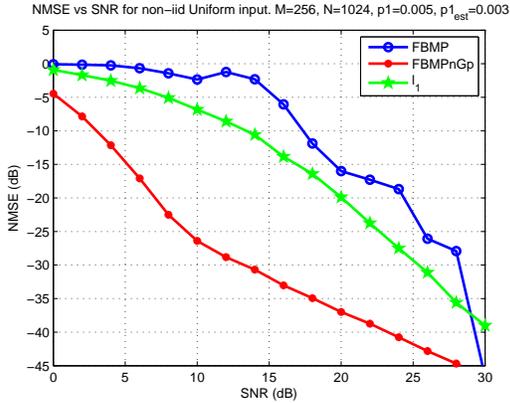}
\label{fig:Uniform_varying_SNR_a}}
\hfil
\subfloat[NMSE vs SNR for Gaussian i.i.d. input]{\includegraphics[scale=0.5] {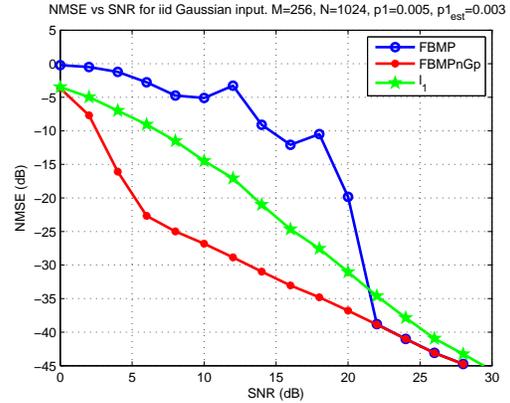}
\label{fig:Gaussian_fixed_SNR_b}}}
\caption{NMSE vs SNR graphs for uniform non-i.i.d. and Gaussian i.i.d. inputs}
\label{fig:Uniform_varying_SNR}
\end{figure*}

Figs. \ref{fig:Uniform_varying_SNR_a} and \ref{fig:Gaussian_fixed_SNR_b} show that the proposed method has better NMSE performance than both FBMP and CS for all considered signals. Only at very high values of SNR does the NMSE performance of FBMP approach that of nGpFBMP.

\subsection*{Experiment 2 (Signal estimation performance comparison for varying sparsity parameter $p$)}
In a similar set of experiments, NMSE and mean runtime were measured for different values of sparsity parameter $p$. The value of SNR selected for these experiments was $20$ dB. 

\begin{figure*}[htbp]
\centerline{\subfloat[NMSE vs $p$]{\includegraphics[scale=0.5]{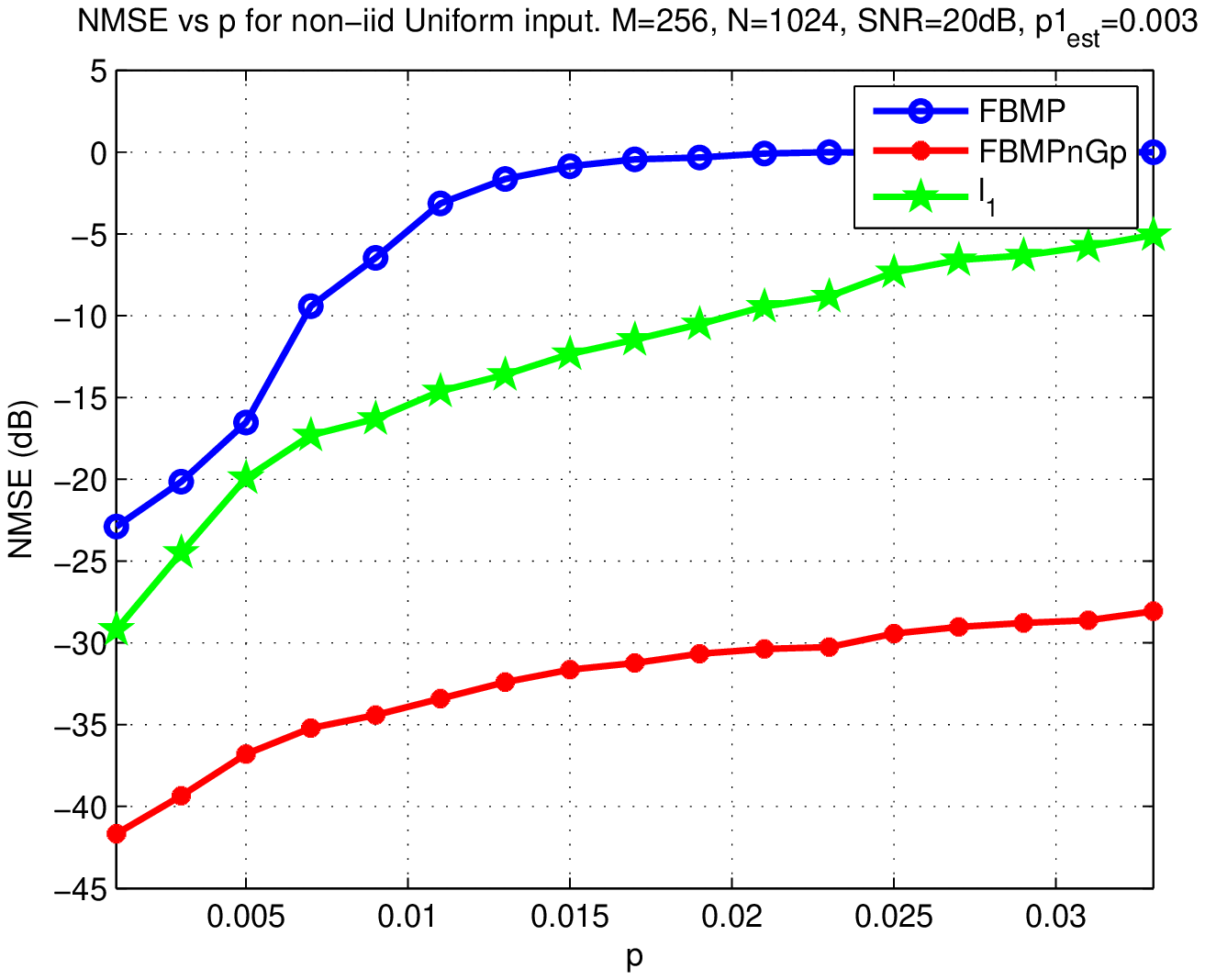}
\label{fig:Uniform_varying_p_a}}
\hfil
\subfloat[Runtime vs $p$]{\includegraphics[scale=0.5] {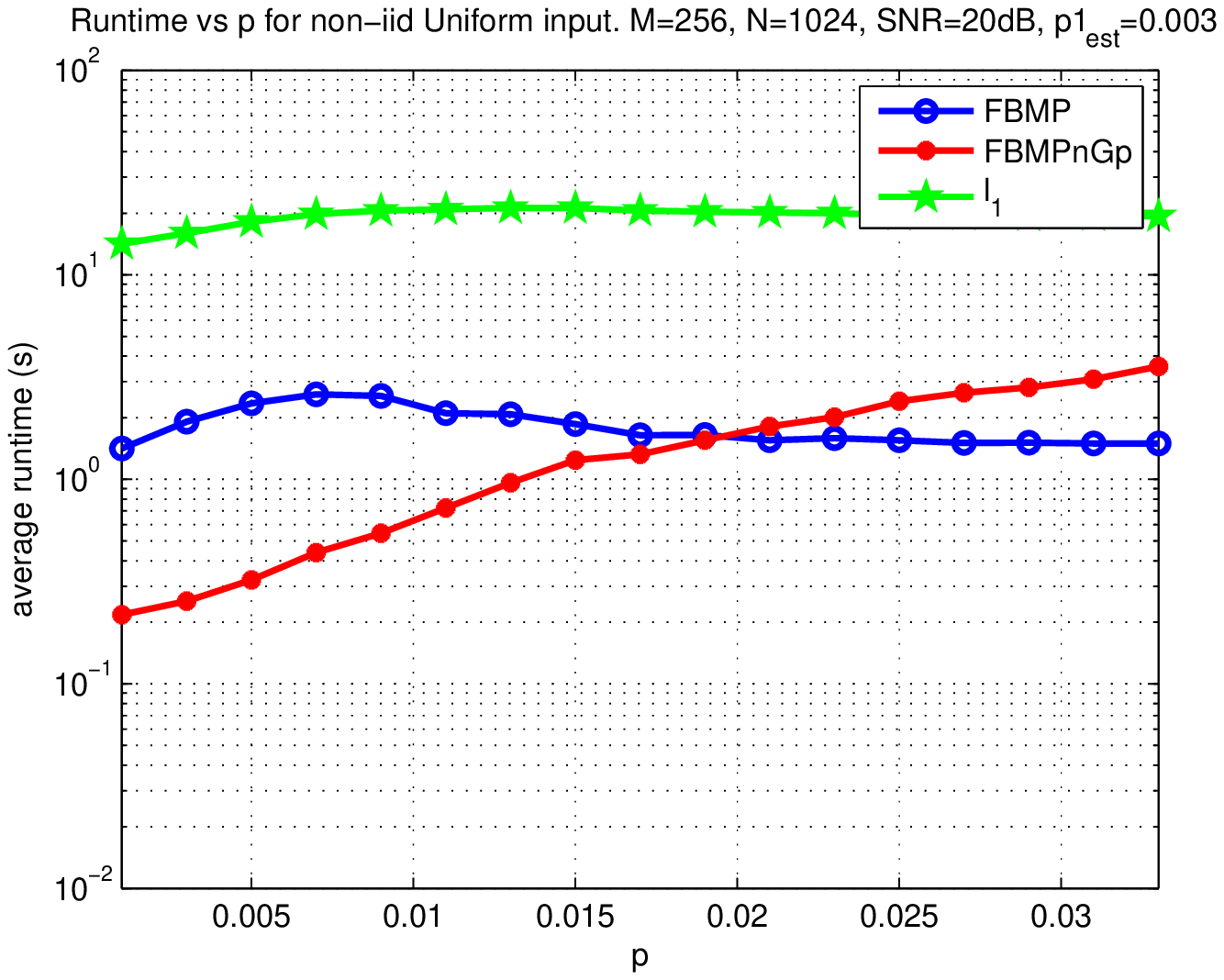}
\label{fig:Uniform_varying_p_b}}}
\caption{NMSE and average runtime vs $p$ graphs for uniform non-i.i.d. input}
\label{fig:Uniform_varying_p}
\end{figure*}

\begin{figure*}[htbp]
\centerline{\subfloat[NMSE vs $p$]{\includegraphics[scale=0.5]{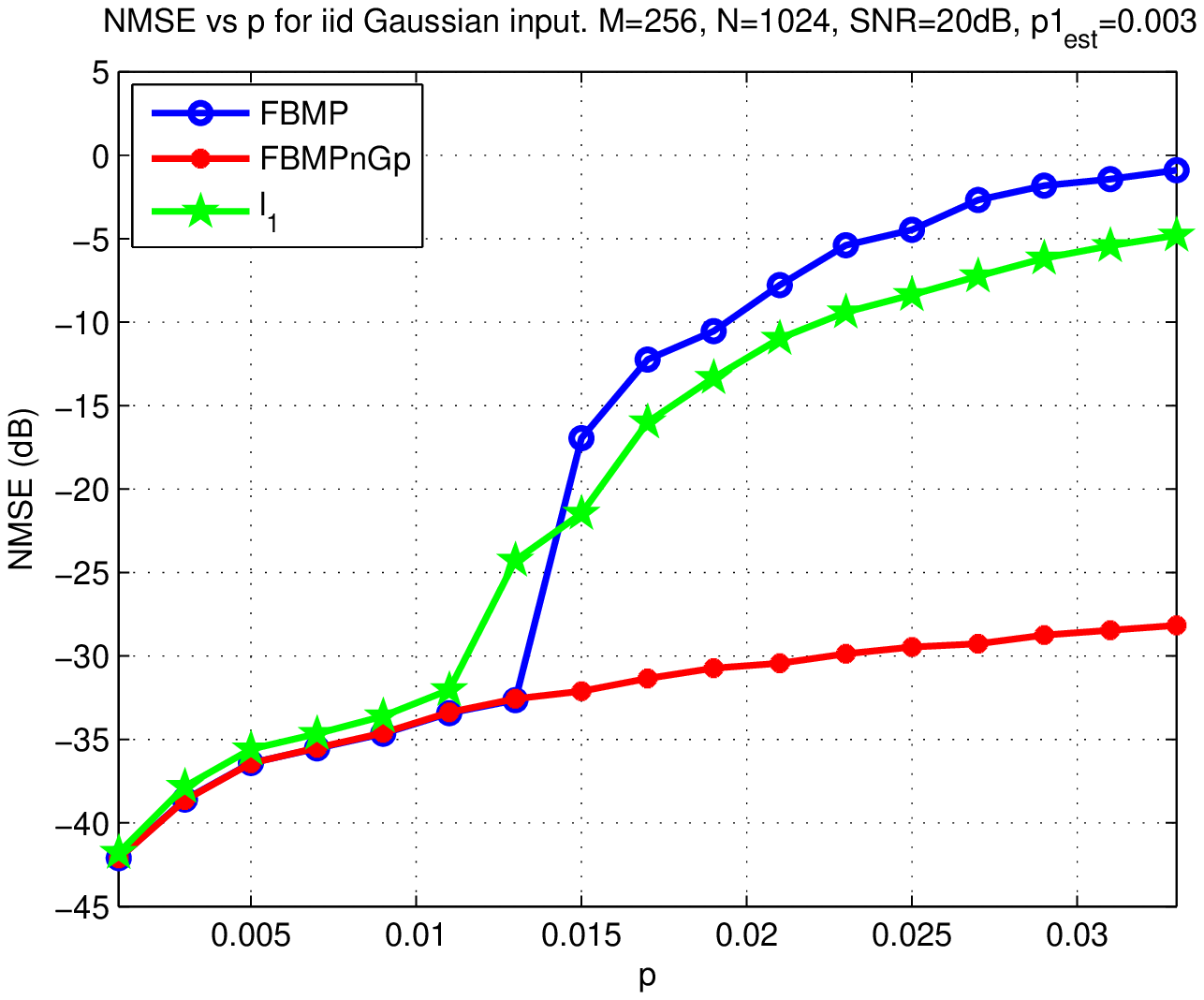}
\label{fig:Gaussian_fixed_p_a}}
\hfil
\subfloat[Runtime vs $p$]{\includegraphics[scale=0.5] {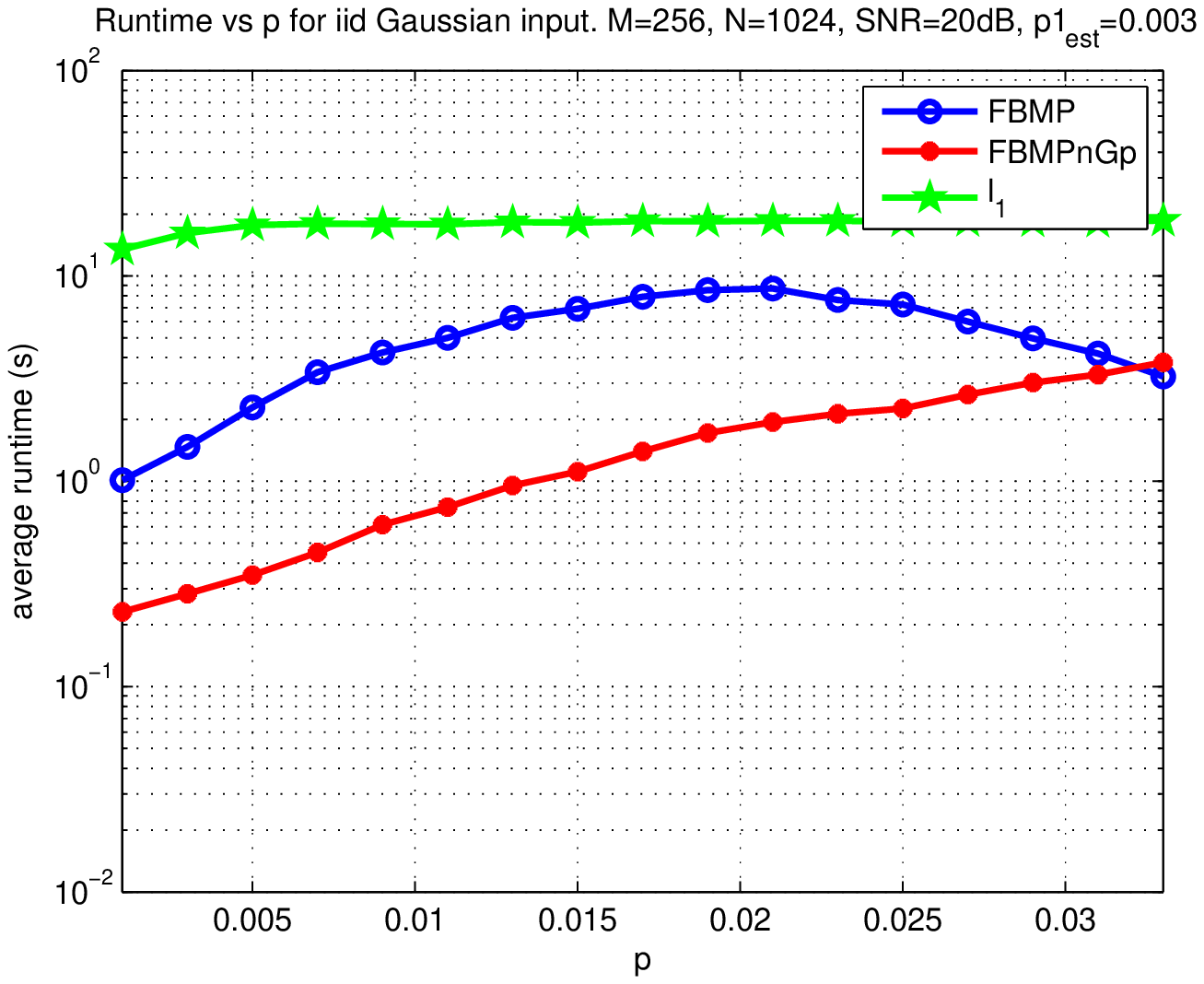}
\label{fig:Gaussian_fixed_p_b}}}
\caption{NMSE and average runtime vs $p$ graphs for Gaussian i.i.d. input}
\label{fig:Gaussian_fixed_p}
\end{figure*}

Figs. \ref{fig:Uniform_varying_p} and \ref{fig:Gaussian_fixed_p} demonstrate the superiority of nGpFBMP over FBMP and CS. Runtime graphs of Figs. \ref{fig:Uniform_varying_p} and \ref{fig:Gaussian_fixed_p} depict that the runtime of nGpFBMP increases for higher values of $p$. This occurs because the initial estimate of $p$ was $0.003$, and as the sparsity rate of $\xv$ increased, more iterations were required to estimate the value of $p$. With higher values of $p$, the difference in runtime is insignificant given the excellent NMSE performance of our method. We also observe that performance of nGpFBMP is relatively insensitive to changes in $p$ as the corresponding changes in NMSE are very small, thus demonstrating the strength of the proposed algorithm.

\subsection*{Experiment 3 (Comparison of signal estimates when the initial statistics of signal and noise are very close to the actual values)}
Table \ref{tab:comparison} compares the average NMSEs of FBMP and nGpFBMP for different types of signals when the initial estimates ($\mu_\xv, \text{ } \sigma_\xv^2, \text{ and }\sigma_\nv^2$) were chosen to be very near to their actual values. Since nGpFBMP is independent of these initial estimates its performance did not change. On the other hand, performance of FBMP improved, although it did not outperform nGpFBMP.
The results show the robustness of our algorithm as it is not dependent on the initial estimates of $\mu_\xv, \text{ } \sigma_\xv^2, \text{ and }\sigma_\nv^2$. We note that each value in Table \ref{tab:comparison} has been computed by averaging the results of 500 independent experiments.

\begin{table}[!t]
\renewcommand{\arraystretch}{1.3}
\caption{Average NMSE comparison between FBMP and nGpFBMP when the initial estimates of the hyperparameters are close to the actual values. Values are in dB.}
\label{tab:comparison}
\centering
\begin{tabular}{|c|c|c|}
\hline
\textbf{Signal type} & \textbf{FBMP} & \textbf{nGpFBMP}\\
\hline
Gaussian & $-20.55$ & $-31.103$\\
\hline
Uniform (i.i.d.) & $-24.2$ & $-30.98$\\
\hline
Uniform (non-i.i.d.) & $-23.87$ & $-30$\\
\hline
\end{tabular}
\end{table}

\subsection*{Experiment 4 (Comparison of multiscale image recovery performance)}

\begin{figure}
	\centering
		\includegraphics[scale=0.5] {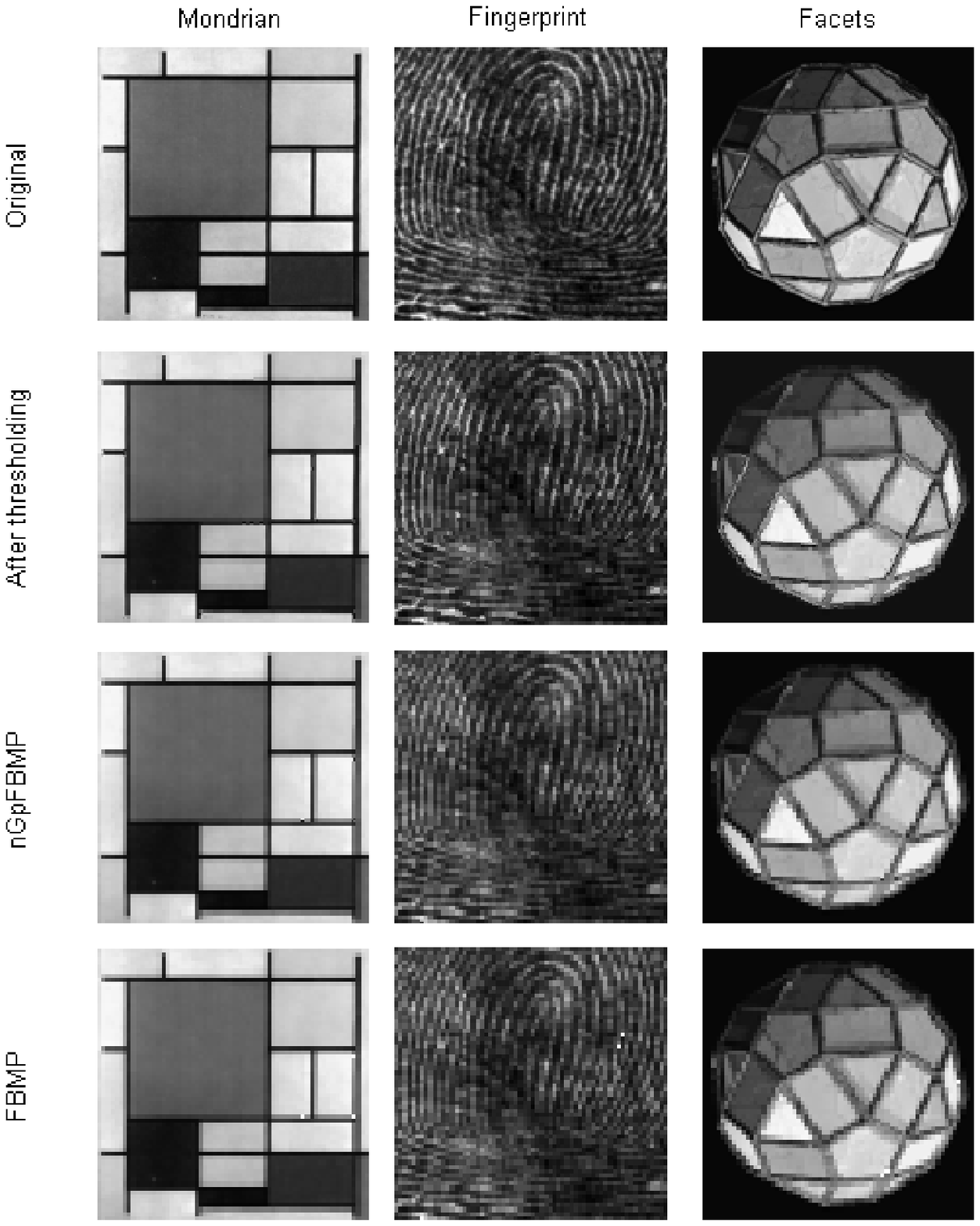}
		\caption{Row 1: Original images used in the multiscale recovery experiment. Row 2: Reconstructed images (inverse wavelet transform) once the wavelet-decomposed detail components were subjected to the threshold. Rows 3 \& 4: Reconstructed from random measurements using nGpFBMP and FBMP respectively.}
	\label{fig:images}
\end{figure}

In another experiment, we carried out multiscale recovery of different images that were $128 \times 128$ pixels. These images are shown in the first row of Fig. \ref{fig:images}. One-level Haar wavelet decomposition of these images was performed, resulting in one `approximation' (low-frequency) and three `detail' (high-frequency) images. Unlike the approximation component, the detail components are compressible in nature. We, therefore, generated their sparse versions by applying a suitable threshold; the noisy random measurements were acquired later from the threshold versions. The detail components were reconstructed from these measurements through nGpFBMP. Finally, inverse wavelet transform was applied to reconstruct the images from the recovered details and the original approximations. Reconstruction errors and times were recorded and, for comparison, recoveries were obtained using FBMP. The resulting reconstructed images are shown in Fig. \ref{fig:images}. Numerical details of the results for these experiments are given in Table \ref{tab:images}, from which it is obvious that images reconstructed using nGpFBMP have lower NMSEs and require a significantly shorter reconstruction time than does FBMP.

\begin{table}[!t]
\renewcommand{\arraystretch}{1.3}
\caption{NMSE ($\mathbf{\epsilon}$) and Reconstruction time comparisons between FBMP and nGpFBMP for the test images shown in Fig. \ref{fig:images}}
\label{tab:images}
\centering
\begin{tabular}{|l|c|c|c|c|}
\hline
\multicolumn{1}{|c|}{\textbf{Image}} & \multicolumn{2}{c|}{\textbf{FBMP}} &  \multicolumn{2}{c|}{\textbf{nGpFBMP}}\\
\cline{2-5}
 & $\mathbf{\epsilon}$ \textbf{(dB)} & \textbf{Time (s)} & $\mathbf{\epsilon}$ \textbf{(dB)} & \textbf{Time (s)} \\
\hline
Mondrian   & $-16.68$  & $19.36$&$-18.84$ & $6.8$ \\\hline
Fingerprint   & $-9.97$  & $19.24$&$-13.92$ & $6.75$ \\\hline
Facets & $-14.75$ & $19.56$&$-19.30$ & $7.01$ \\
\hline
\end{tabular}
\end{table}

\section{Conclusion}\label{sec:conclusions}
In this paper, we presented a robust Bayesian matching pursuit algorithm based on a fast recursive method. Compared with other robust algorithms, our algorithm does not require signals to be derived from some known distribution. This is useful when we can not estimate the parameters of the signal distributions. Application of the proposed method on several different signal types demonstrated its superiority and robustness.

\bibliographystyle{IEEEtran}
\bibliography{savedrecs}

\section*{Appendix}
\subsection*{Algorithm: Non-Gaussian prior fast Bayesian matching pursuit}\label{app:pseudocode}
\begin{table}[h!]
\label{alg:nGpFBMP}
\begin{enumerate}
	\item The following quantities are calculated one time and used in the following iterations
	\begin{enumerate}
		\item\label{itm:ydotA} inner products of $\yv$ with columns of $\Phim$.
		\item\label{itm:AcoldotAcol} inner products of columns of $\Phim$ with each other.
		\item $\mathbf {e}_{\yv, 1}(i) = \left(\boldsymbol \phi_{i}^\herm \boldsymbol \phi_{i}\right)^{-1} \boldsymbol \phi_{i}^\herm	\yv$ for $i=1, 2, \dots, N$ i.e. $\mathbb{E}[\xv|\yv,\Sc]$ for each of the one-length support vector. Note that the components of these calculations are available from steps \ref{itm:ydotA} and \ref{itm:AcoldotAcol}.
	\end{enumerate}
	\item Calculate $N$ one-element metrics, $\nu(\Sc_1)$, using $\mathbf {e}_{\yv, 1}$ from the previous step for each of the possible support elements $\alpha_i,  i\in[1,N]$ as explained in Section \ref{sec:greedy}.
	\item Perform the following search $D$ times (each search has $p=1:P$ stages)
	\begin{enumerate}
		\item At each $p$th stage: \label{itm:p}
		\begin{enumerate}
			\item\label{itm:startp} Find the $p$-element metric, $\nu(\Sc_p)$, with the maximum value and note down the corresponding $\alpha_i$.
			\item If this $\alpha_i$ has been selected in any of the previous $p$th stages, mark it and discard it and go to step \ref{itm:startp} again.
			\item else update $\mathbf {e}_{\boldsymbol \phi, p}$ using (\ref{eq:e_recurs}) for this newly selected position, $\alpha_i$, and add the corresponding $\Sc_p$ as a new member of $\Sc^d$. \label{itm:ephi}
			\item Save $\mathbf {e}_{\boldsymbol \phi, p}$ to update its value in the next iteration.
			\item Use $\mathbf {e}_{\boldsymbol \phi, p}$ and $\mathbf {e}_{\yv, p}$ to compute the update $\mathbf {e}_{\mathbf y, p+1}$  (\ref{eq:ey_recur}) for all $\Sc_{p+1} = [\Sc_p \quad \alpha_i],  \forall i \mid \alpha_i \notin \Sc_p$.\label{itm:ey}
			\item Compute the dominant support selection metrics for all these combinations $\Sc_{p+1}$ using the computed $\mathbf {e}_{\yv, p+1}$. This yields ($p+1$)-element metrics $\nu(\Sc_{p+1})$ to be used in the next iteration.\label{itm:endp}
		\end{enumerate}
		\item end $p$
	\end{enumerate}
	\item $\Sc^d$ contains the dominant supports that will be used to find the sum in (\ref{eq:xammse}).
\end{enumerate}
\end{table}

\end{document}